# A numerical analysis of the impact of gas pressure and dielectric material on the generation of body force in an air gas plasma actuator

Sajad Hajikhani[a], Ramin Mehrabifard[b,c], Hamed Soltani Ahmadi[a,d]

[a] Department of Atomic and Molecular Physics, Faculty of Science, University of Mazandaran, Babolsar, Iran

[b] Department of Physics and Institute for Plasma Research, Kharazmi University, 49 Dr. Mofatteh Avenue, Tehran, Iran

[c] Division of Environmental Physics, Faculty of Mathematics, Physics and Informatics, Comenius University, Mlynská dolina, 84248 Bratislava, Slovakia

[d] Plasma Technology Research Core, Faculty of Science, University of Mazandaran, Babolsar, Iran

**Abstract**

Plasma technology has undeniably revolutionized industrial processes in recent decades. Atmospheric pressure plasma (APP) has emerged as a prominent and widely applicable tool in various scientific disciplines. Notably, plasma-assisted flow control has become a subject of intense interest, particularly applying surface dielectric barrier discharge (SDBD) plasma actuators for aerodynamic flow control. In this study, a two-dimensional model of the SDBD plasma actuator is developed using the COMSOL Multiphysics program, incorporating air gas discharge reactions with $N_2/O_2/Ar$ gases in specific ratios (0.78, 0.21, 0.01). The investigation focuses on the impact of dielectric materials (mica, silica glass, quartz, and polytetrafluoroethylene (PTFE)) on plasma characteristics and body force within the plasma actuator under constant input parameters. Moreover, the study explores how variable pressure (760, 660, and 560 torr) in different applications influences plasma properties, ultimately affecting the magnitude of the body force in the plasma actuator. These findings contribute to optimizing plasma technology for flow control applications and enhance industrial efficiency and performance.

# 1-Introduction

Since nonthermal plasma first attracted scientific and technical attention 70 years ago, plasma technology has taken center stage in industrial processes both now and in the future. Over time, plasma technology developed and is today employed in everything from everyday items to cutting-edge applications (Keidar & Beilis, 2013). Various scientific sectors have utilized APP because of its outstanding quality (Assadi et al., 2021; Da Ponte et al., 2011; Mehrabifard et al., 2017, 2020; Venezia et al., 2008). Over the past few years, plasma-assisted flow control has received much interest (Neretti et al., 2014; Science & 2008, n.d.). Significant interest is in utilizing surface dielectric barrier discharge (SDBD) plasma actuators to control aerodynamic flow. Plasma actuators find utility in the realm of active airfoil leading edge separation control (Neretti et al., 2014), used for high lift (Little et al., 2010), boundary layer flow control (Porter et al., 2007; Szulga et al., 2015), handling dynamic stall in an airfoil (Post & Corke, 2006), bluff body flow control (Do et al., 2007), regulation of airflow (Neretti et al., 2012), lowering noise levels (Thomas et al., 2008), and postponing turbine blade separation (Huang et al., 2006). Many studies have been conducted experimentally and numerically, focusing on optimizing the ionic wind velocity and the volumetric force generation mechanism (Jayaraman & Shyy, 2008; A. V. Likhanskii et al., 2008; Mahdavi & Sohbatzadeh, 2019; Moreau, 2007). Moreau et al. reviewed the mechanical and electrical characteristics and their applications in aerodynamic flow control (Benard & Moreau, 2014). Numerous earlier numerical research has utilized two or more straightforward reactions to reduce the computations required (Abdollahzadeh et al., 2012; Boeuf et al., 2007; A. A. V Likhanskii et al., 2007). Additionally, the plasma component may occasionally be viewed under an electrostatic condition (Abdelraouf et al., 2020; Kazemi et al., 2021; Omidi & Mazaheri, 2020; Tehrani et al., 2022; Yu et al., 2023). In some cases, a specific sort of gas has also been applied to this structure (Mehrabifard, 2023).

This study describes a two-dimensional model of the SDBD plasma actuator. The COMSOL Multiphysics program is used in the development of the model. Air gas discharge reactions with the combination of nitrogen, oxygen, and argon gas with a ratio of 0.78, 0.21, and 0.01, respectively, are considered for this simulation. With constant input parameters, we investigate the effect of dielectric material on the plasma characteristics and body force in the plasma actuator. Besides that, the pressure can be variable in many of the mentioned applications, which can change many parameters of the plasma, which, as a result, changes the magnitude of the body force in the plasma actuator.

## 2- Model Description

### 2-1- Governing Equations

In this investigation, the fluid model was employed. The electron density and energy can be calculated by resolving the drift-diffusion formulas. To formulate the governing equations of electric discharge, the drift-diffusion approximation was adopted (Mehrabifard, 2023):

$$\frac{\partial n_e}{\partial t} + \nabla \cdot \vec{\Gamma}_e = R_e - (\vec{u} \cdot \nabla) n_e \tag{1}$$

$$\vec{\Gamma}_e = -(\vec{\mu}_e \cdot \vec{E}) n_e - \vec{D}_e \cdot \nabla n_e \tag{2}$$

The electron continuity equation is defined by equation (1). $n_e$ the electron density, $D_e$ diffusion coefficient, $\Gamma_e$ electron flux, $u$ average species velocity, and $R_e$ electron generation rate are all given in equation (1). Equation (2) represents the electron flow, divided into drift and diffusion. The electron energy density can be calculated using this equation:

$$\frac{\partial n_\varepsilon}{\partial t} + \nabla \cdot \vec{\Gamma}_\varepsilon + \vec{E} \cdot \vec{\Gamma}_\varepsilon = R_\varepsilon - (\vec{u} \cdot \nabla) n_\varepsilon \tag{3}$$

$$\vec{\Gamma}_\varepsilon = -(\vec{\mu}_\varepsilon \cdot \vec{E}) \cdot n_\varepsilon - \vec{D}_\varepsilon \cdot \nabla n_\varepsilon$$

The value of $\vec{E} \cdot \vec{\Gamma}_\varepsilon$ is the amount of energy that can be extracted from an electron by applying an electric field. The following equation may be used to compute the energy gained by non-elastic collisions, which is denoted by the variable $R_\varepsilon$:

$$R_\varepsilon = S_{en} + \frac{Q + Q_{gen}}{q} \tag{4}$$

$S_{en}$ is the power dissipation, $Q_{gen}$ is the heat source, and q is the electron charge. $D_e$ is electron diffusion coefficient, $\mu_\varepsilon$ indicates energy mobility, and $D_\varepsilon$ is energy distribution coefficient. The link between these parameters is shown in Equation 5:

$$D_\varepsilon = \mu_\varepsilon T \quad D_e = \mu_e T_e \quad \mu_\varepsilon = \frac{5}{3} \mu_e \tag{5}$$

The Townsend coefficients of the electron source, which are determined by the following equation, were used:

$$R_e = \sum_{j=1}^{M} x_j a_j N_n |\Gamma_e| \tag{6}$$

Where $M$ is the total number of reactions, $x_j$ the molar fraction of the target species for reaction j, $a_j$ the Townsend coefficient, and $N_n$ the total number of neutral particles are present. Considering the number p of non-elastic electron collisions, we will have:

$$R_\varepsilon = \sum_{j=1}^{p} x_j a_j N_n |\Gamma_e| \Delta\varepsilon_j \tag{7}$$

Which $\Delta\varepsilon_j$ is the energy dissipation of the j reaction. For non-electron-induced species, the below equation is used for mass fraction calculation:

$$\rho \frac{\partial w_k}{\partial t} + \rho(\vec{u}.\nabla) w_k = \nabla.\vec{j}_k + R_k \tag{8}$$

In which, $w_k$ is the ionic density $j_k$ is the energy flux of the ions. The following equation obtains the electrostatic field:

$$\nabla.(\varepsilon_0 \varepsilon_r E) = \rho \tag{9}$$

Where, $\varepsilon_0$ is the permittivity of vacuum, and $\varepsilon_r$ is a relative dielectric constant. The following relationships are found regarding the boundary conditions for the electron flux and energy flow:

$$-\hat{n}.\vec{\Gamma}_e = (\frac{1}{2} v_{eth} n_e) - \sum_p \gamma_p (\vec{\Gamma}_p.\hat{n}) \tag{10}$$

$$-\hat{n}.\vec{\Gamma}_\varepsilon = (\frac{5}{6} v_{eth} n_e) - \sum_p \varepsilon_p \gamma_p (\vec{\Gamma}_p.\hat{n}) \tag{11}$$

The right-hand side of equation 10 displays the electron number caused by the secondary electron and $\gamma$ represents the secondary electron coefficient. Ions and excited species on the surface of electrodes are neutralized via the surface reaction. Surface interactions on the electrode are indicated by the $\beta_j$ coefficient, which means the probability of the function of the j species. The definition of flux matching for each heavy species is as follows:

$$n.j_k = M_k R_{surf,k} + M_k c_k \mu_{m,k^z k}(n.E)[(z_k n.E) > 0] \tag{12}$$

In which, $j_k$ and $R_{surf,k}$ indicate the diffusive flux vector and the surface reaction rate expression for species k. $M_k$ is mass fraction and $c_k$ is particle mass density.

The body force generated by plasma is:

$$f_{coulomb} = \rho_v \times normE \qquad (13)$$

In this equation, $\rho_v$ is the density of electrons and positive/negative ions, $normE$ is the normalized electric field in the x-y direction. In this simulation, the value of charge density is equal to:

$$\rho_v = (n_{Ar+} + n_{N_2+} + n_{O_2+} + n_{N_4+} + n_{O_+} + n_{O_{4+}} - n_e - n_{O-}) \qquad (14)$$

### 2-2- Boundary Conditions and Reactions

An asymmetric pair of copper electrodes separated by a dielectric substance from the actuator. While the other electrode is connected to the ground and encased by the dielectric material, the first electrode is placed on the dielectric surface and in contact with the gas flow at atmospheric pressure. A Gaussian voltage power supply drives the discharge at 1.5 kV. The electron density is first estimated to be $10^8$ m$^{-3}$. The gas is at a temperature of 293.15 K and a pressure of 760 torr. In the simulation, the material components are defined as a model of the air, considering the related reactions of the species. Rate coefficients were obtained by solving Boltzmann's equation with BOLSIG+ (Hagelaar & Pitchford, 2005) and the cross-sections from the LXCAT data source (Pitchford et al., 2017). Reaction rates were taken from the references (Sakiyama et al., 2012). The dominant reactions can be seen in Tables 1 to 6, including electron impact ionization, electron attachment, elastic collisions, excitation, recombination, neutral component collisions, and ion conversion processes. The schematic structure of SDBD and its boundary conditions is shown in Figure 1. Copper electrode dimensions are ($0.05mm \times 5mm$), and the dielectric dimensions are ($0.5mm \times 10mm$). We employ five boundary layers and a 1.4 stretching factor for the system's whole boundary while meshing this structure. A free triangular mesh with a maximum element size of 0.4 is employed for the entire geometry (Figure 2).

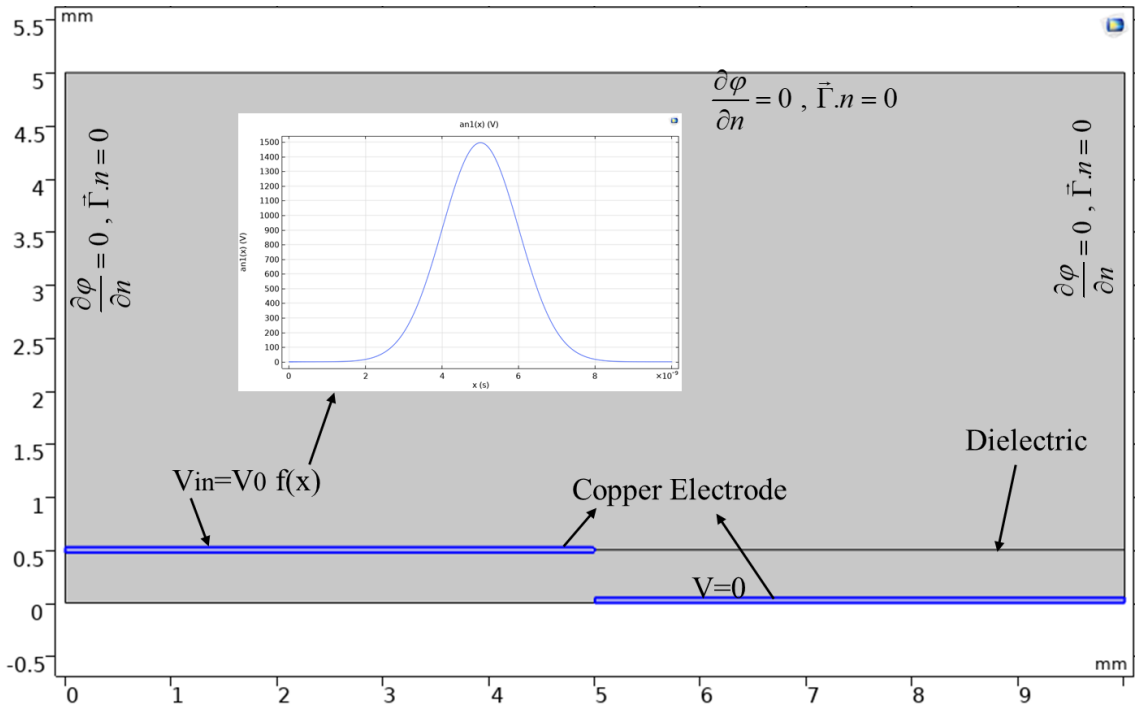

Figure 1. Schematic of the plasma actuator, boundary condition, and input voltage with its function

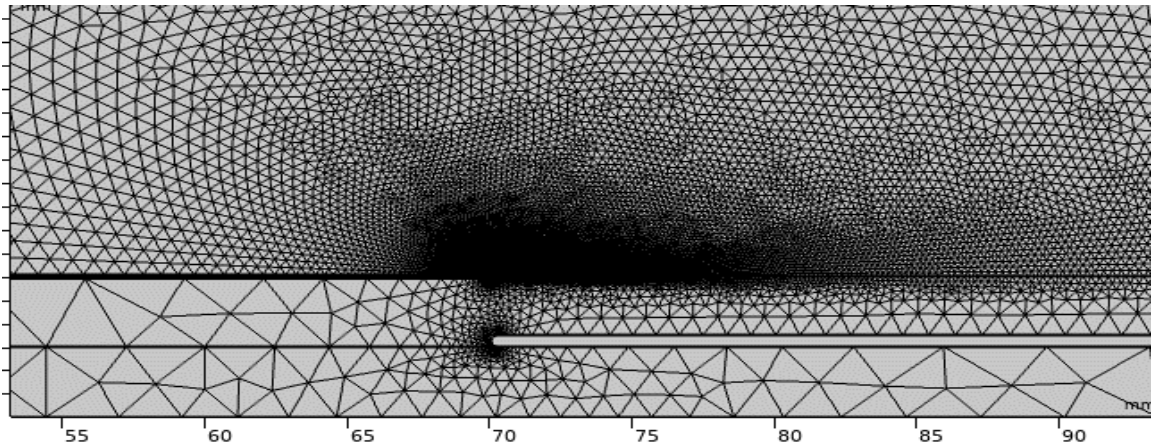

Figure 2. Meshing and its density in edges

Nitrogen, oxygen, and argon reactions with a mass fraction of 0.78, 0.21, and 0.01 are used to simulate air gas discharge. The electron impact reactions with active nitrogen species, such as nitrogen molecules $N_2$ and $N_2^+$, that are singly ionized are shown in Table 1. The interactions of electron impact with active oxygen species, such as metastable oxygen $O_2s$, oxygen molecule $O_2$, and singly ionized oxygen molecule $O_2^+$, are shown in Table 2. Tables 3,4 and 5 offer the two and three-body reaction rates between atoms and molecules. Table 6 shows all electron impact reactions with argon and its molecule-to-molecule reactions. Fifteen surface reactions are considered in addition to the aforementioned reactions, as indicated in Table 7.

Table1. The reactions of electron impact with active species of nitrogen. (Singly ionized nitrogen molecule $N_2^+$, metastable nitrogen $N_2s$, nitrogen molecule $N_2$) (*BOLSIG+ / Electron Boltzmann Equation Solver*, n.d.; Mehrabifard, 2023; University of Toulouse, n.d.)

| Reactions | Formula | Type | $\Delta\varepsilon(eV)$ |
|---|---|---|---|
| 1 | $e + N_2 \rightarrow e + N_2$ | Elastic | 0 |
| 2 | $e + N_2 \rightarrow e + N_2s$ | Excitation | 0.02 |
| 3 | $e + N_2 \rightarrow e + N_2s$ | Excitation | 0.29 |
| 4 | $e + N_2 \rightarrow e + N_2s$ | Excitation | 0.291 |
| 5 | $e + N_2 \rightarrow e + N_2s$ | Excitation | 0.59 |
| 6 | $e + N_2 \rightarrow e + N_2s$ | Excitation | 0.88 |
| 7 | $e + N_2 \rightarrow e + N_2s$ | Excitation | 1.17 |
| 8 | $e + N_2 \rightarrow e + N_2s$ | Excitation | 1.47 |
| 9 | $e + N_2 \rightarrow e + N_2s$ | Excitation | 1.76 |
| 10 | $e + N_2 \rightarrow e + N_2s$ | Excitation | 2.06 |
| 11 | $e + N_2 \rightarrow e + N_2s$ | Excitation | 2.35 |
| 12 | $e + N_2 \rightarrow e + N_2s$ | Excitation | 6.17 |
| 13 | $e + N_2 \rightarrow e + N_2s$ | Excitation | 7 |
| 14 | $e + N_2 \rightarrow e + N_2s$ | Excitation | 7.35 |
| 15 | $e + N_2 \rightarrow e + N_2s$ | Excitation | 7.36 |
| 16 | $e + N_2 \rightarrow e + N_2s$ | Excitation | 7.8 |
| 17 | $e + N_2 \rightarrow e + N_2s$ | Excitation | 8.4 |
| 18 | $e + N_2 \rightarrow e + N_2s$ | Excitation | 8.16 |
| 19 | $e + N_2 \rightarrow e + N_2s$ | Excitation | 8.55 |
| 20 | $e + N_2 \rightarrow e + N_2s$ | Excitation | 8.89 |
| 21 | $e + N_2 \rightarrow e + N_2s$ | Excitation | 11.03 |
| 22 | $e + N_2 \rightarrow e + N_2s$ | Excitation | 11.88 |
| 23 | $e + N_2 \rightarrow e + N_2s$ | Excitation | 12.25 |
| 24 | $e + N_2 \rightarrow e + N_2s$ | Excitation | 13 |

| | 25 | $e + N_2 \rightarrow 2e + N_2^+$ | ionization | 15.6 |
|---|---|---|---|---|

Table 2. The interactions between oxygen and electrons (*BOLSIG+ / Electron Boltzmann Equation Solver*, n.d.; University of Toulouse, n.d.)

| Reactions | Formula | Type | $\Delta\varepsilon(eV)$ |
|---|---|---|---|
| 1 | $e + O_2 \rightarrow e + O_2$ | Elastic | - |
| 2 | $e + O_2 \rightarrow O + O^-$ | Attachment | - |
| 3 | $e + O_2 \rightarrow e + O_2$ | Excitation | 0.02 |
| 4 | $e + O_2 \rightarrow e + O_2$ | Excitation | 0.19 |
| 5 | $e + O_2 \rightarrow e + O_2$ | Excitation | 0.19 |
| 6 | $e + O_2 \rightarrow e + O_2$ | Excitation | 0.38 |
| 7 | $e + O_2 \rightarrow e + O_2$ | Excitation | 0.38 |
| 8 | $e + O_2 \rightarrow e + O_2$ | Excitation | 0.57 |
| 9 | $e + O_2 \rightarrow e + O_2$ | Excitation | 0.75 |
| 10 | $e + O_2 \rightarrow e + O_2a1d$ | Excitation | 0.977 |
| 11 | $e + O_2a1d \rightarrow e + O_2$ | Excitation | -0.977 |
| 12 | $e + O_2 \rightarrow e + O_2b1s$ | Excitation | 1.627 |
| 13 | $e + O_2b1s \rightarrow e + O_2$ | Excitation | -1.627 |
| 14 | $e + O_2 \rightarrow e + O_245$ | Excitation | 4.5 |
| 15 | $e + O_245 \rightarrow e + O_2$ | Excitation | -4.5 |
| 16 | $e + O_2 \rightarrow e + O + O$ | Dissociation | 6 |
| 17 | $e + O_2 \rightarrow e + O + O1d$ | Excitation | 8.4 |
| 18 | $e + O_2 \rightarrow e + O + O1s$ | Excitation | 9.95 |
| 19 | $e + O_2 \rightarrow 2e + O_2^+$ | Ionization | 12.06 |

Table 3. Atomic and molecule-to-molecule interactions with two and three bodies (Sohbatzadeh & Soltani, 2018; Stafford & Kushner, 2004)

| Reactions | Formula | Type | $K^f$ (m$^3$/s.mol) |
|---|---|---|---|
| 1 | $e + N^+ \rightarrow N$ | Recombination | $3.5 \times 10^{-18}$ |
| 2 | $e + N_2 \rightarrow 2e + N + N^+$ | Dissociative ionization | $2.4 \times 10^{-23}$ |
| 3 | $e + N_2 \rightarrow e + 2N$ | Dissociative | $2 \times 10^{-17}$ |
| 4 | $e + N_2^+ \rightarrow 2N$ | - | $2.8 \times 10^{-13}$ |

| | | |
|---|---|---|
| 5 | $N^+ + N_2 \rightarrow N + N_2^+$   Charge exchange | $10^{-17}$ |

Table 4. Atomic and molecule-to-molecule interactions with two and three bodies (Sohbatzadeh & Soltani, 2018; Stafford & Kushner, 2004)

| Reactions | Formula | $K^f$ (m$^3$/s.mol) |
|---|---|---|
| 1 | $O + O_2 + O_2 \rightarrow O_3 + O_2$ | $6 \times 10^{-46} \times (1.3^{-2.8})$ |
| 2 | $O + O_2 + N_2 \rightarrow O_3 + N_2$ | $5.6 \times 10^{-46} \times (1.3^{-2.8})$ |
| 3 | $O + O_3 \rightarrow O_2 + O_2$ | $8 \times 10^{-18} \times \exp(-2060/4)$ |
| 4 | $O + NO_2 \rightarrow NO + O_2$ | $5.6 \times 10^{-7} \times \exp(180/40)$ |
| 5 | $O + NO_3 \rightarrow O_2 + NO_2$ | $1.7 \times 10^{-17}$ |
| 6 | $O + N_2O_5 \rightarrow NO_2 + NO_2 + O_2$ | $1 \times 10^{-22}$ |
| 7 | $N + O_2 \rightarrow NO + O$ | $1.5 \times 10^{-7} \times \exp(8)$ |
| 8 | $N + O_3 \rightarrow N + O_2$ | $1 \times 10^{-22}$ |
| 9 | $N + NO \rightarrow N_2 + O$ | $2.1 \times 10^{-11} \times \exp(0.25)$ |
| 10 | $NO + O_3 \rightarrow NO_2 + O_2$ | $3 \times 10^{-18} \times \exp(-3.8)$ |
| 11 | $N + NO_2 \rightarrow N_2O + O$ | $5.8 \times 10^{-18} \times \exp(0.55)$ |
| 12 | $NO_2 + O_3 \rightarrow NO_3 + O_2$ | $1.4 \times 10^{-19} \times \exp(-6.2)$ |
| 13 | $O_2b1s + O_2 \rightarrow O_3 + O$ | $4.8 \times 10^{-21}$ |
| 14 | $N_2 + O_2 \rightarrow N_2O + O$ | $6 \times 10^{-20} \times (1.3^{0.55})$ |
| 15 | $O^- + O \rightarrow O_2 + e$ | $2 \times 10^{-16} \times (400^{0.5})$ |
| 16 | $O^- + O_2 \rightarrow O_3 + e$ | $3 \times 10^{-16} \times (400^{0.5})$ |
| 17 | $O^- + O_2 \rightarrow O + O_2 + e$ | $6.9 \times 10^{-16} \times (400^{0.5})$ |
| 18 | $O_2 + O_2 \rightarrow O_3 + O$ | $2.95 \times 10^{-27} \times (400^{0.5})$ |
| 19 | $O^- + O_3 \rightarrow O_2 + O_2 + e$ | $3 \times 10^{-16} \times (400^{0.5})$ |
| 20 | $O^- + O_3 \rightarrow O_2^- + O_2$ | $1.02 \times 10^{-17} \times (400^{0.5})$ |
| 21 | $O_2^- + O \rightarrow O_3 + e$ | $1.5 \times 10{-16} \times (400^{0.5})$ |
| 22 | $O + O_3 \rightarrow O_2 + O + O$ | $1.2 \times 10^{-16}$ |
| 23 | $O_2 \rightarrow O_2$ | $0.2$ |
| 24 | $O_2 \rightarrow O$ | $10^{-5}$ |
| 25 | $N^+ + O_2 \rightarrow N + O_2^+$ | $3 \times 10^{-16}$ |

Table 5. Atomic and molecule-to-molecule interactions with two and three bodies (Sohbatzadeh & Soltani, 2018)

| Reactions | Formula | $K^f$ (m$^6$/s.mol$^2$) |
|---|---|---|
| 1 | $N_2 + Ar + N_2^+ \rightarrow Ar + N_4^+$ | $1.8 \times 10^6$ |
| 2 | $e + N_4^+ \rightarrow N_2 + N_2$ | $1.2 \times 10^{11}$ |
| 3 | $O_2 + Ar + O_2^+ \rightarrow Ar + O_4^+$ | $2 \times 10^5$ |
| 4 | $O_4^+ + Ar + O^- \rightarrow 2O_2 + Ar + O$ | $5.2 \times 10^{11}$ |
| 5 | $Ar + O_4^+ \rightarrow Ar + O + O_2 + e + O^+$ | $6 \times 10^7$ |
| 6 | $O^- + Ar + O_4^+ \rightarrow 2O_2 + O + Ar$ | $3.8 \times 10^8$ |
| 7 | $N_2 + Ar + N_2^+ \rightarrow Ar + N_4^+$ | $1.8 \times 10^6$ |

Table 6. The interactions between electrons and Argon (*BOLSIG+ / Electron Boltzmann Equation Solver*, n.d.; University of Toulouse, n.d.)

| Reactions | Formula | Type | $\Delta\varepsilon(eV)$ |
|---|---|---|---|
| 1 | $e + Ar \rightarrow e + Ar$ | Elastic | 0 |
| 2 | $e + Ar \rightarrow e + Ars$ | Excitation | 11.5 |
| 3 | $e + Ars \rightarrow e + Ar$ | Superelastic | -11.5 |
| 4 | $e + Ar \rightarrow 2e + Ar^+$ | Ionization | 15.8 |
| 5 | $e + Ars \rightarrow 2e + Ar^+$ | Ionization | 4.24 |
| 6 | $Ars + Ars \rightarrow e + Ar + Ar^+$ | Penning ionization | - |
| 7 | $Ars + Ar \rightarrow Ar + Ar$ | Metastable quenching | - |

Table 7. Table of surface reactions in air discharge (Sohbatzadeh & Soltani, 2018)

| Reactions | Formula | Sticking Coefficient |
|---|---|---|
| 1 | $O_2a1d \rightarrow O_2$ | 1 |
| 2 | $O_245 \rightarrow O_2$ | 1 |
| 3 | $O_2b1s \rightarrow O_2$ | 1 |
| 4 | $O_2 \rightarrow O_2$ | 1 |
| 5 | $O_2^+ \rightarrow O_2$ | 1 |
| 6 | $O^+ \rightarrow O$ | 1 |
| 7 | $O^- \rightarrow O$ | 1 |
| 8 | $O1s \rightarrow O$ | 1 |
| 9 | $O1d \rightarrow O$ | 1 |

| 10 | $N^+ \rightarrow N$ | 1 |
| 11 | $Ars \rightarrow Ar$ | 1 |
| 12 | $Ar^+ \rightarrow Ar$ | 1 |
| 13 | $N_2 s \rightarrow N_2$ | 1 |
| 14 | $Ns \rightarrow N$ | 1 |
| 15 | $N_2^+ \rightarrow N_2$ | 1 |

## 3- Results and Discussion

An SDBD plasma actuator's body force magnitude has been obtained from the simulations. The rate of species generation is significantly influenced by the dielectric substance. Dielectric materials are electrically insulating substances with low electrical conductivity. Numerous processes that alter the behavior of the plasma and the emergence of species may take place when it comes into contact with a dielectric substance (Fridman et al., 2016; Lieberman & Lichtenberg, 2005).

In this study, we investigated how four different dielectric materials affected plasma properties that change the magnitude of body force in the plasma actuator. Figure 3 displays the electrical potential when plasma is formed. The grounded electrode has zero potential, while the power electrode has a voltage of 1.5 kV with a Gaussian shape.

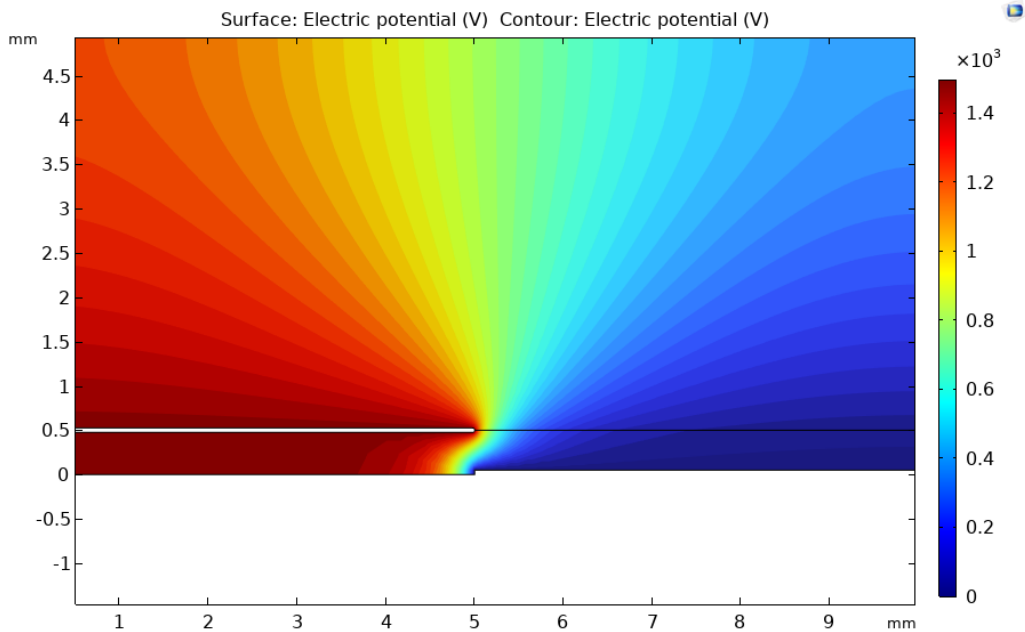

Figure 3. Distribution of electric potential for $V_{in}=1.5$ (*KV*)

Electron density and temperature play a vital role in plasma and are the initiators of many plasma reactions. And in many applications, the effect of electron density has been investigated (Mehrabifard et al., 2023; Tanaka et al., 2015), and it is an effective factor for the plasma actuators. The evaluation of electron temperature at 6 ns for various dielectric materials is shown in Figure 4. Moving away from the power electrode causes the electron temperature to decrease from its highest value. As the figure shows, the temperature changes for each dielectric material were almost in the same range. Changing materials does not make a significant difference in electron temperature. The temperature of the electron is directly influenced by the electric field due to the implementation of a constant potential function, resulting in minimal fluctuations in temperature.

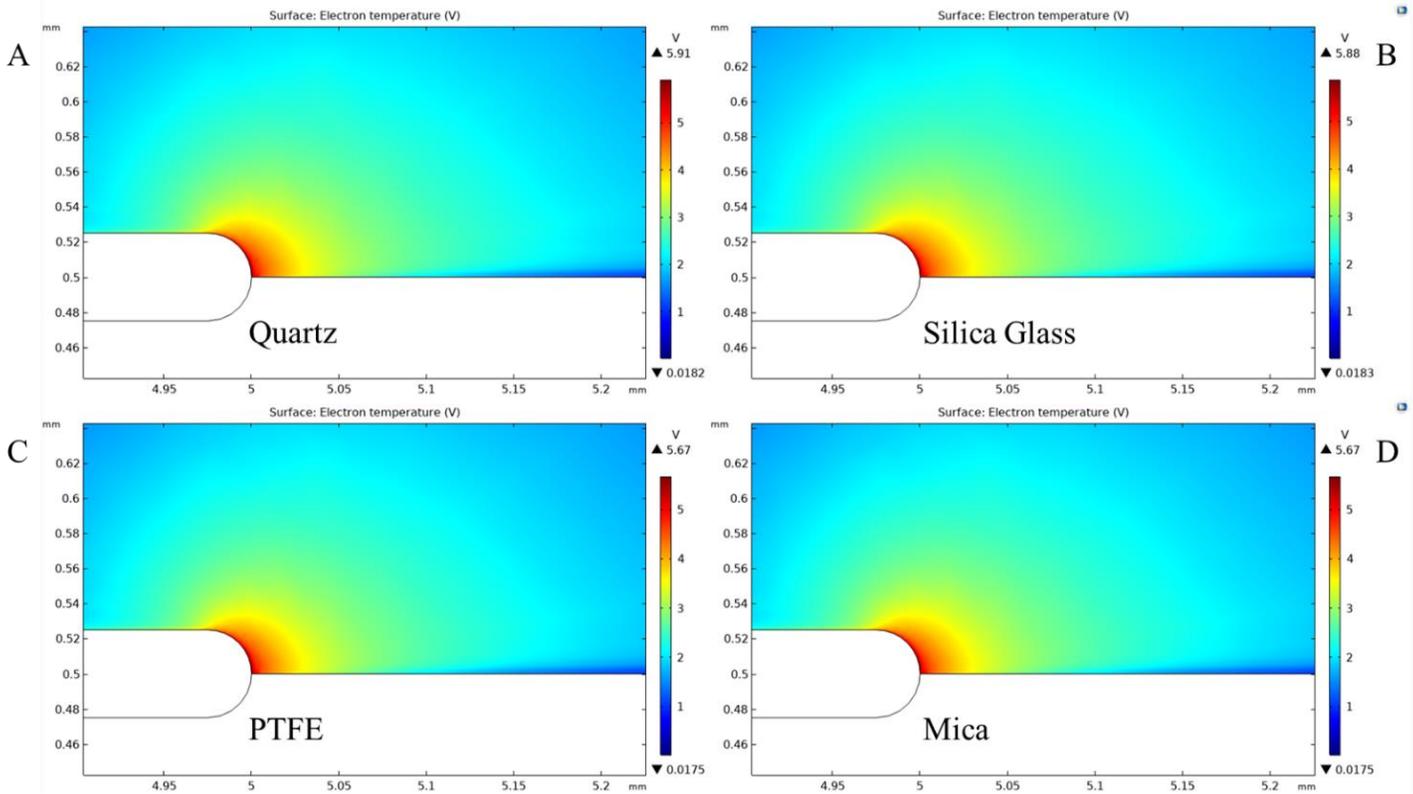

Figure 4. The electron temperature over 6 *ns* at 1.5 kV with Gaussian function for different dielectric materials (A) Quartz, (B) Silica Glass, (C) PTFE, (D) Mica

Then, once more, Equations 12 and 13 say this: the magnitude of the body force is directly influenced by the particle density. When all other factors remain the same, figure 5 illustrates how much the electron density changes with respect to the dielectric material. Using mica as dielectric results in electrons in its maximum value; quartz and silica have almost the same amount, $8.61 \times 10^{14} \, (1/m^3)$ and $8.22 \times 10^{14} \, (1/m^3)$ respectively, and PTFE has the lowest value of $2.39 \times 10^{14} \, (1/m^3)$. Changes in electron density are influenced by many causes, including surface charge accumulation and the photoionization effect. Among these factors, the alteration of the dielectric coefficient is particularly significant in determining the changes in surface charge accumulation. Indeed, the density rises in proportion to the increase in the dielectric coefficient.

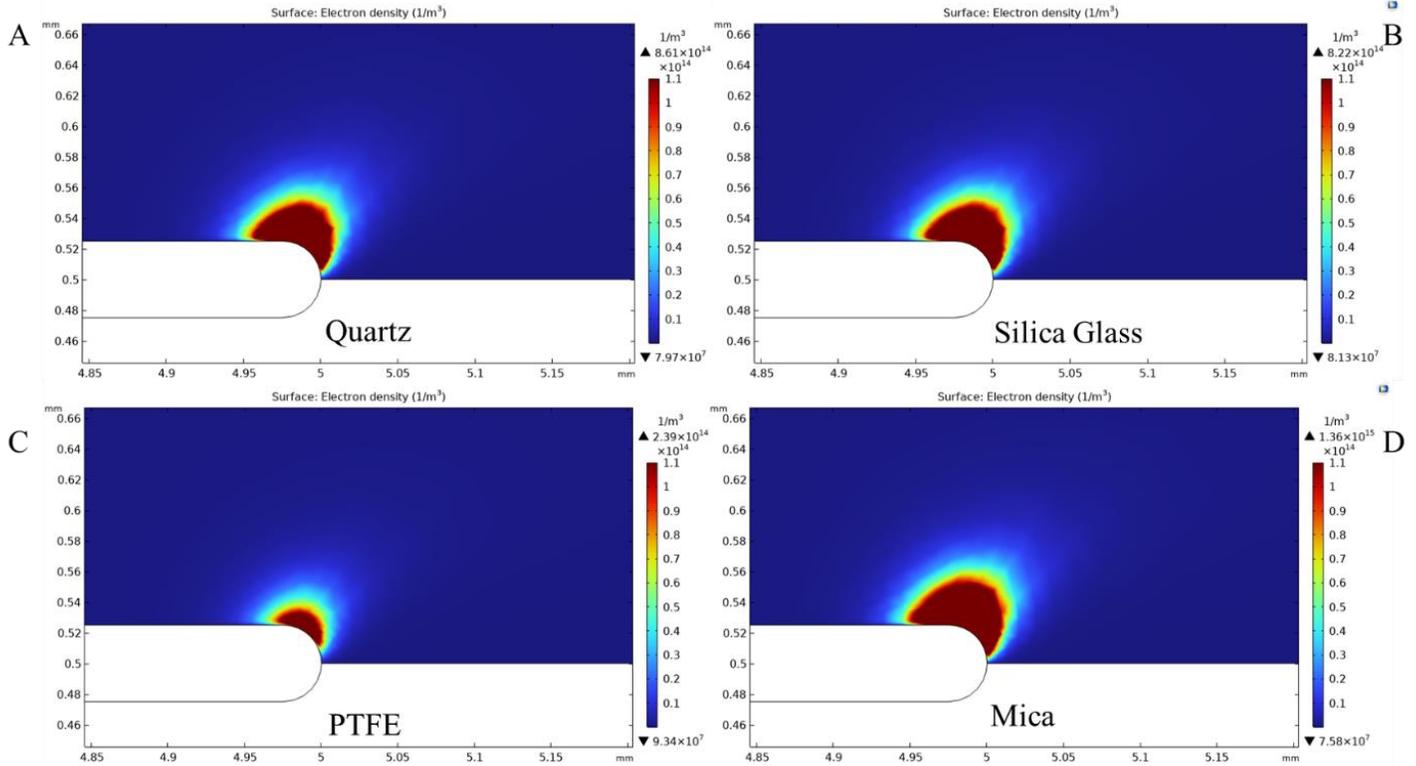

Figure 5. The electron densities over 6 *ns* at 1.5kV with Gaussian shape for different dielectric materials (A) Quartz, (B) Silica Glass, (C) PTFE, (D) Mica

The ion densities are shown in Figure 6. The $n_{Ar+}, n_{N_2+}, n_{O_2+}, n_{N_4+}, n_{O_+}, n_{O_{4+}}, n_{O-}$ are the main species measured in this simulation in the presence of different dielectric materials. As Figure 6 shows, all ion densities have a higher value in the presence of mica and the lowest for PTFE. And the value for quartz and silica glass shows almost the same ion density. As stated for the electron. The increase in surface charge accumulation will affect the density of ions, which is related to the dielectric coefficient.

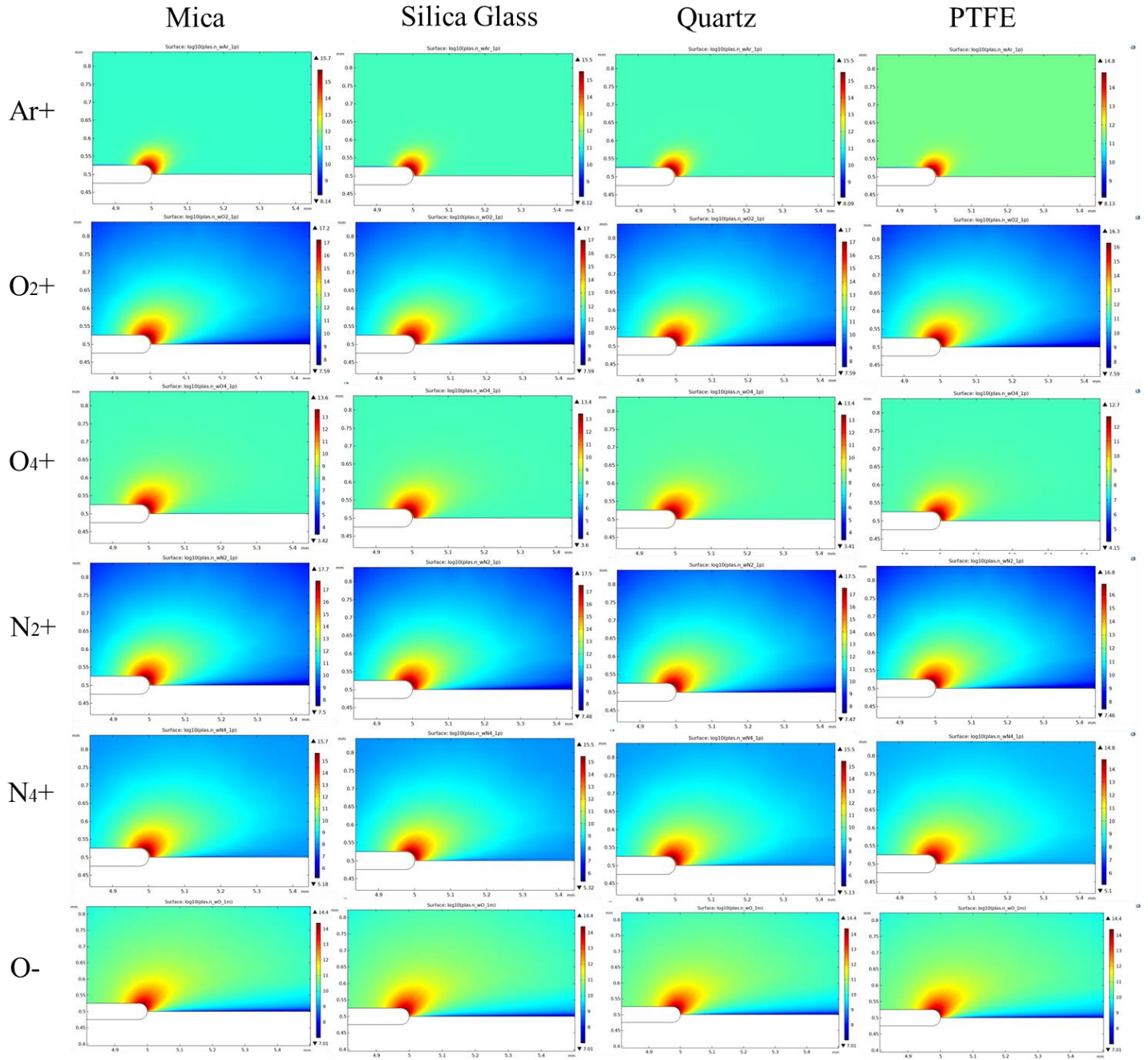

Figure 6. Two-dimensional distribution of the ion densities for different dielectric materials

The body force can be calculated from the difference between positive and negative charges and the magnitude of the normal field in the discharge space. According to Equation 13, the logarithmic body force distribution for dielectric materials is shown in figure 7. To investigate more precisely, a hypothetical line on the surface of the electrode and dielectric is considered. This line's beginning and end points are (x=4, y=0.54) and (x=8, y=0.54). Figure 8 shows a non-logarithmic magnitude of the body force on this virtual line. As it is known, the force of the body is the highest for Mica, and its magnitude will be equal to 9800 (N/m$^3$). It will be equal to 5700 (N/m$^3$), 5600 (N/m$^3$), and 1100 (N/m$^3$) for quartz, silica, and PTFE, respectively.

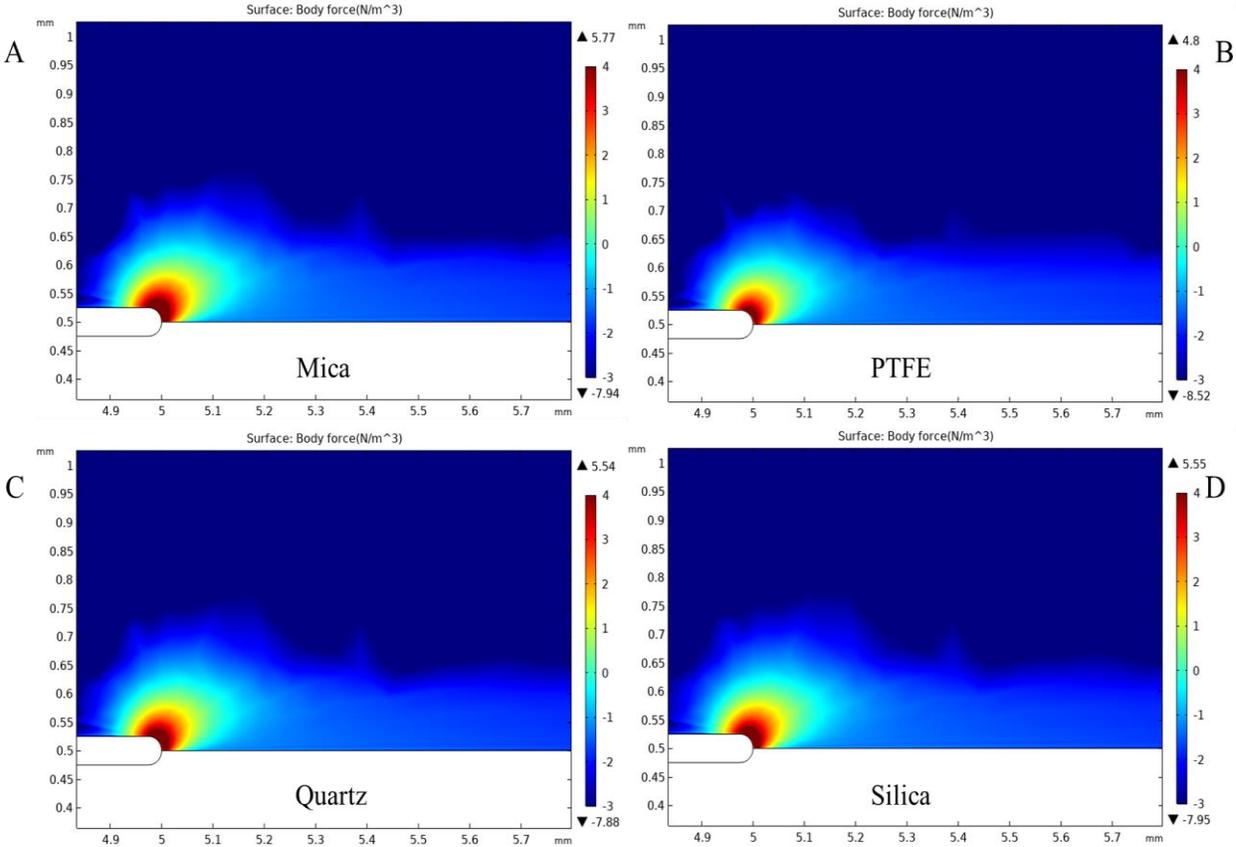

Figure 7. Magnitude of body force in the presence of different dielectric barrier discharge

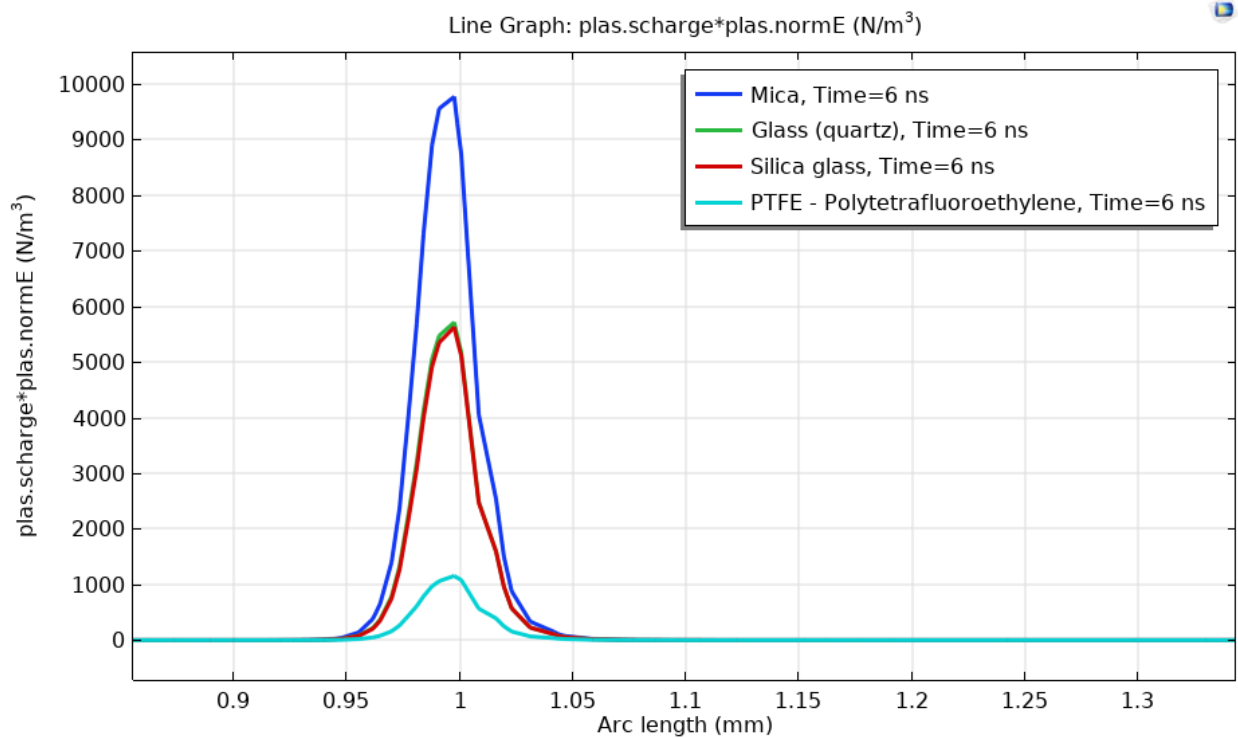

Figure 8. The body force is created on the surface of the electrode and the dielectric.

Gas pressure is one of the main parameters that can change the characteristics of the plasma; as a result, this causes a change in the magnitude of the body force in the plasma actuators. The pressure of 560, 660, and 760 torr are considered for plasma simulation. Figure 9 shows changes in body force magnitude for different pressures. The body force is measured on a virtual line on the upper part of the power electrode. As shown in the figure, the magnitude of body force changes dramatically by reducing pressure. The pressure was reduced by 100 torr in each stage, but significant changes were shown in body force.

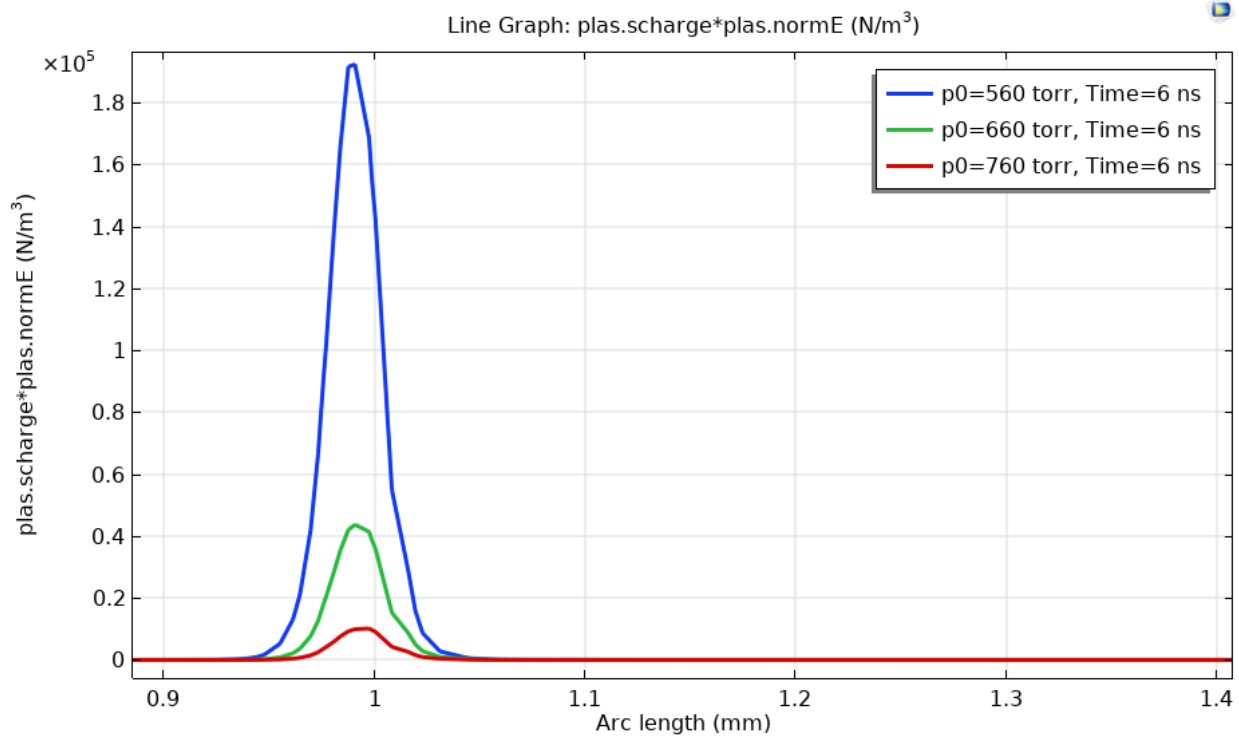

Figure 9. The effect of different pressures on body force

Changing the dielectric coefficient, the electron energy distribution function changes around the power electrode, and this change leads to different ionization collisions around this electrode, which ultimately creates a difference in the ionization coefficient and volumetric force. As the ambient pressure changes, the ratio of the electric field to the ambient pressure will change, considering that the drift velocity is a function of the electric field-to-pressure ratio (E/P). On the other hand, this velocity is directly related to electron mobility; it can affect the changes in electron mobility. In fact, the set of these net charge density changes that lead to the production of propulsion force will be different

**4-Conclusion**

Active current control is one of the areas where plasma is used. Additionally, modeling these systems before construction may save time and money and ensure the creation of an efficient system. From the outcome of our investigation, it is possible to conclude that without changing the main parameters and only by changing the type of dielectric material, the magnitude of the body force can be increased. Among the selected materials, using mica, we have the most body force. Furthermore, it is worth noting that pressure, being a crucial factor in plasma production, has shown that even the slightest alteration may lead to substantial variations in the body's force. Moreover, due to the significant change in body force, pressure is one of the parameters that should be considered in the actuator design. The simulation findings, performed with consideration of air gas discharge, can be valuable in the design of plasma drive systems for many applications, enabling the attainment of optimum outcomes via the selection of appropriate materials.